\documentclass[final,3p,times]{elsarticle}

\usepackage{amssymb}
\usepackage{amsmath}
\usepackage{moreverb,url}
\usepackage[colorlinks,bookmarksopen,bookmarksnumbered,citecolor=red,urlcolor=red]{hyperref}
\usepackage[capitalise,noabbrev]{cleveref}
\usepackage{caption,subcaption}
\usepackage{orcidlink}
\usepackage[export]{adjustbox}
\usepackage{booktabs}
\usepackage{tabularx}
\usepackage{bm}
\usepackage{soul}
\usepackage{multirow}
\usepackage{multicol}

\journal{\href{https://www.mdpi.com/journal/vibration}{Vibration}}
\begin{document}
\begin{frontmatter}



\title{Damping Identification Sensitivity in Flutter Speed Estimation\tnoteref{label5}}
\tnotetext[label5]{\hl{This is an author-accepted manuscript version of the following open access article: G. Dessena, A. Pontillo, M. Civera, D. I. Ignatyev, J. F. Whidborne, and L. Zanotti Fragonara, ‘Damping identification sensitivity in flutter speed estimation,’ \emph{Vibration}, vol. 8, no. 2, p. 24, May 2025. doi: 10.3390/vibration8020024. [Online]. Available:} [\url{https://doi.org/10.3390/vibration8020024}]}

\author[label1,label2]{Gabriele Dessena \orcidlink{0000-0001-7394-9303}}
\affiliation[label1]{organization={School of Aerospace, Transport and Manufacturing, Cranfield University},
            addressline={College Road},
            city={Cranfield},
            postcode={MK43 0AL},
            state={England},
            country={UK}}

\affiliation[label2]{organization={Department of Aerospace Engineering, Universidad Carlos III de Madrid},
            addressline={Av.da de la Universidad 30},
            postcode={28911},
            city={Leganés},
            state={Madrid},
            country={Spain}}

\author[label3]{Alessandro Pontillo \orcidlink{0000-0003-0015-685X}} 

\affiliation[label3]{organization={School of Engineering, UWE Bristol},
            addressline={Frenchay Campus, Coldharbour Lane}, 
            city={Bristol},
            postcode={BS16 1QY}, 
            state={England},
            country={UK}}

\author[label4]{Marco Civera \orcidlink{0000-0003-0414-7440}} 

\affiliation[label4]{organization={Department of Structural, Geotechnical and Building Engineering, Politecnico di Torino},
            addressline={Corso Duca degli Abruzzi 24}, 
            city={Turin},
            postcode={10129}, 
            state={Piedmont},
            country={Italy}}
            
\author[label1]{Dmitry I. Ignatyev \orcidlink{0000-0003-3627-3740}} 
\author[label1]{James F. Whidborne \orcidlink{0000-0002-6310-8946}} 
\author[label1]{Luca Zanotti Fragonara \orcidlink{0000-0001-6269-5280}} 
\begin{abstract}
Predicting flutter remains a key challenge in aeroelastic research, with certain models relying on modal parameters, such as natural frequencies and damping ratios. These models are particularly useful in early design stages or for the development of small Unmanned Aerial Vehicles (maximum take-off mass below 7 kg). This study evaluates two frequency-domain system identification methods, Fast Relaxed Vector Fitting (FRVF) and the Loewner Framework (LF), for predicting the flutter onset speed of a flexible wing model. Both methods are applied to extract modal parameters from Ground Vibration Testing data, which are subsequently used to develop a reduced-order model with two degrees of freedom. Results indicate that FRVF and LF-informed models provide reliable flutter speed, with predictions deviating by no more than 3\% (FRVF) and 5\% (LF) from the N4SID-informed benchmark. The findings highlight the sensitivity of flutter speed predictions to damping ratio identification accuracy and demonstrate the potential of these methods as computationally efficient alternatives for preliminary aeroelastic assessments.
\end{abstract}



\end{frontmatter}


\section{Introduction}

The accurate characterisation of aircraft dynamic behaviour is critical to ensure structural safety and performance optimisation under various operational conditions. The extraction of modal parameters -- natural frequencies ($\omega_n$), damping ratios ($\zeta_n$), and mode shapes ($\bm{\phi}_n$) -- via system identification (SI) techniques, allows characterisation of the structural dynamics of the system, such that it can be used for finite element model (FEM) updating and subsequent aeroelastic analyses. Experimental Modal Analysis (EMA), or Ground Vibration Testing (GVT) in the aeronautics domain, provides the necessary framework to identify these parameters from experimental data. These techniques are pivotal for understanding how dynamic systems behave, offering insights that drive critical applications, such as flutter prediction \cite{VanTran2023}, structural model updating \cite{Yuan2025}, and vibration-based damage detection \cite{Dessena2022}.

{While SI techniques are well-established for modal parameter identification, their application to complex, large-scale aerostructures presents several challenges.}
{Frequency-domain system identification is widely used in modal analysis; however, certain fitting processes – especially those relying on direct parametric and rational fraction polynomial curve-fitting} \cite{Champneys2023,Haywood-Alexander2024} {-- can exhibit ill-conditioning when dealing with noisy or highly complex datasets} \cite{Civera2021}. 
{However, several advanced frequency-domain techniques, such as the pLSCF estimator} \cite{Guillaume2003} {have shown excellent numerical stability and robustness even in these challenging conditions} \cite{El-Kafafy2013,Steffensen2025}.

Similarly, time-domain techniques such as Numerical Algorithms for Subspace State Space System Identification (N4SID) \cite{VanOverschee2005} are computationally demanding and face scalability issues when processing extensive input-output data \cite{Dessena2022g}. {In terms of scalability, this happens to other time-domain Stochastic Subspace Identification (SSI) techniques, such as SSI with Canonical Variate Analysis, which struggled -- only few modes out of 42 accurately identified -- in the identification of high-dimensional (5 input and 21 output channels) aeronautical system }\cite{Dessena2024c}. Recent advances in SI, including methods such as the Fast Relaxed Vector Fitting (FRVF) \cite{Civera2021a} and the Loewner Framework (LF) \cite{Dessena2022}, have sought to address these limitations by improving robustness and computational efficiency. These innovations have shown significant promise in EMA and GVT, providing a pathway for broader application for aeroelastic onset speed estimation.

Flutter analysis is a crucial aspect of aeroelastic research, representing a significant challenge at the intersection of aero- and structural engineering. Flutter occurs due to dynamic instability arising from aeroelastic coupling between aerodynamic forces and structural modes, potentially leading to catastrophic failures \cite{Dowell1969, Edwards1983}. In aeronautics, flutter prediction aims to determine the critical velocity at which oscillatory instabilities occur. Established techniques, such as modal damping extrapolation \cite{Dimitriadis2001}, time-marching computational fluid dynamics \cite{Chen2007}, and semi-analytical methods like the $p-k$ approach \cite{Zhao2012}, have been widely used for this purpose. Computational advancements have enhanced these approaches, improving their predictive accuracy and relevance \cite{Chen2007}. {These advances have fostered the development of integrated workflows for inexpensive flutter testing platforms, such as that in }\cite{Sanmugadas2024}.

In structural engineering, similar principles are applied to the aerodynamic analysis of cable-supported bridges \cite{Namini1992}. Both fields face shared challenges, including structural flexibility, large aspect ratios, and aeroelastic modes coupling. For instance, system identification techniques initially developed for aerostructures have been adapted to study bridge flutter in \cite{Ge2000}. Despite progress, flutter prediction remains computationally intensive, especially for nonlinear systems or structures with significant modal coupling. Efforts to address this include the development of Reduced Order Models (ROMs) \cite{Chen2007}, data-driven techniques, and hybrid approaches that combine physics-based and experimental data \cite{Tian2024a}.

This work applies frequency-domain SI methods for preliminary flutter speed estimation of a flexible wing structure via the classical $p-k$ method and a simplified ROM. Using experimental data and results from an existing GVT campaign \cite{Dessena2022g}, the study evaluates the sensitivity of the {GVT-}identified parameters, $\omega_n$ and $\zeta_n$, for the development of the ROM for critical aeroelastic speeds prediction. In particular, $\zeta_n$ estimated in wind-off GVTs are considered to be dependent {solely from structural effects} for the sake of aeroelastic modelling, as widely accepted in the community \cite{Wright2014}.  Specifically, it examines the effect of $\zeta_n$ -- widely recognised as the most troublesome modal parameter to identify -- on flutter onset speed prediction on an experimental wing model{, the well-known eXperimental BeaRDS-2 (XB-2) case study} \cite{Pontillo2020}. Modal parameters identified with N4SID are used as a benchmark for those identified via LF and FRVF. Thus, the contributions of this work include:
{The main contributions of this work are:}
\begin{itemize}
    \item {First application of FRVF- and LF-identified modal parameters for a two-degree-of-freedom (2-DoF) aeroelastic model;}
    \item {Development of a reduced-order model for the XB-2 wing;}
    \item {Use of LF and FRVF results for flutter speed estimation;}
    \item {Demonstration of a simplified ROM-based approach for critical speed prediction;}
    \item {Comparative analysis of flutter speed predictions using different identification methods.}
\end{itemize}
{The rest of this paper is structured as follows: Section 2 introduces the experimental setup, identification methods, and ROM formulation. Section 3 presents and discusses the results. Conclusions are drawn in Section 4.}

\section{Materials and Methods\label{sec:met}}
In this section, the overall strategy used in this work for studying the damping identification sensitivity on flutter speed is outlined. The overall workflow is shown in \cref{fig:wfl} and outlined below:

\begin{figure}[!ht]
\centering
    \includegraphics[width=.85\textwidth, keepaspectratio]{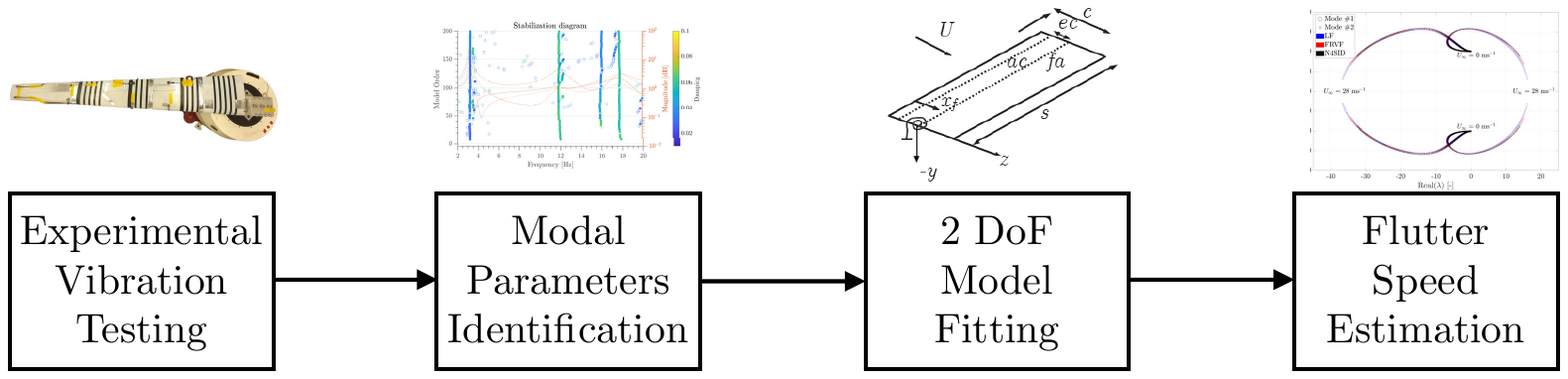}
    \caption{{Flutter speed estimation workflow.}\label{fig:wfl}}
\end{figure}    

\begin{enumerate}
    \item \textbf{Experimental Vibration Testing.} A specimen wing is selected to undergo a GVT campaign, potentially also with different configurations in terms of shape and mass. The specimen and the testing regime are introduced in \cref{sec:xb2}.
    \item \textbf{Modal Parameters Identification.} From the recorded data, modal parameters are identified using different methods. It is not uncommon that this will result in different identified values of $\zeta_n$ across the methods. Here, the LF, FRVF, and N4SID are considered as identification methods; however, N4SID is not discussed in depth as it is a classical and well-known technique. On the other hand, the FRVF and LF are introduced in \cref{sec:FRVF,sec:LF}, respectively.
    \item \textbf{2 DoF Model Fitting.} The modal parameters obtained are used for model fitting. The model itself is described in \cref{sec:aero}.
    \item \textbf{Flutter Speed Estimation.} The classical $p-k$ method is used to obtain the estimated flutter speeds for the different scenarios and methods, thus, giving an idea of the relationship between $\zeta_n$ and the flutter speed itself. These findings are presented in \cref{sec:conc}
\end{enumerate}
Point 1 and the modal parameter identification part of point 2 are covered in greater detail in \cite{Dessena2022g} and are only briefly touched upon in the remainder of this section.

\subsection{The Flexible Wing Model\label{sec:xb2}}
\begin{figure}[!ht]
\centering
\includegraphics[width=.8\textwidth]{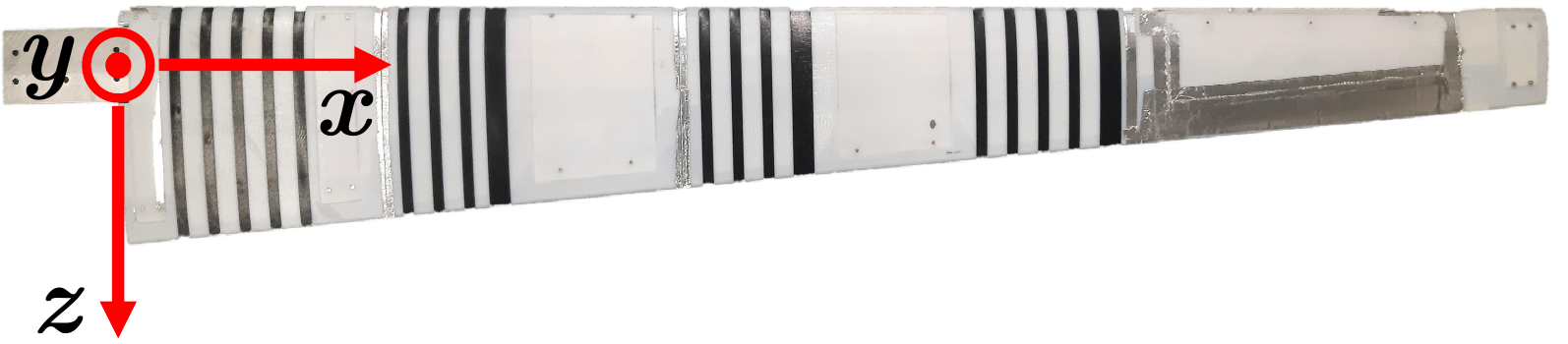}
\caption{XB-2 wing top view. Y-axis coming out of the page. {Any apparent distortions are due to lens-induced optical effects.} (Adapted from \cite{Dessena2023}).
\label{fig:xb2}}
\end{figure}

The experimental case study considered here is the XB-2 high aspect ratio wing (\cref{fig:xb2}) developed within the Beam Reduction and Dynamic Scaling (BeaRDS) project at Cranfield University \citep{Pontillo2018,Yusuf2019,Hayes2019,Pontillo2020}. The XB-2 wing was conceived as a dynamically scaled example of a civil jet airliner wing to be tested in the university wind tunnel. The wing comprises three components: the spar (6082-T6 aluminium), the stiffening tube (stainless steel), and the skin, which is responsible for transferring the aerodynamic loads to the underlying structure and is made of two 3D printed plastics: rigid Digital ABS and rubber-like compound, Agilus 30, visible respectively as the white and black sections in \cref{fig:xb2}. Originally, additional brass masses were used to aid the scaling of mass properties; however, for the purpose of this work, the masses are removed.

The aerodynamic surface of the wing, outlined by a NACA 23015 aerofoil, has {an aerodynamic span ($b_w$) of 1.385 m (1.5 m including the clamp)}, with a mean aerodynamic chord ($\bar{c}$) of 0.172 m, a taper ratio (${\lambda}$) of 0.35, a leading edge (LE) sweep of 1.49\textsuperscript{o}, and a mass of 3.024 kg. The wing has a neutral twist and dihedral angles. 

The wing torque box consists of the spar and tube assembly. The spar was machined from two 6082-T6 aluminium blocks, which were welded together and secured with four bolted L-profile plates \citep{Dessena2022b}. The main spar features a Saint George cross-shaped cross-section and a variable taper along its span, while the tube is a simple stainless steel tube with a 10 mm diameter and 1 mm thickness.

In this work, the GVT results obtained for the scope of damage detection in \cite{Dessena2022g} are considered. To simulate damage, the wing was loaded at different locations and with different mass values, resulting in four scenarios, as shown in \cref{tab:xb2}. This can also simulate different loading scenarios on the wing, such as under-wing payloads.

\begin{table}[!hbt]
\caption{\label{tab:xb2} XB-2 wing loading scenarios. Distances, in mm, are from the origin.}
\centering
\begin{tabularx}{\textwidth}{lXl}
\hline
\textbf{Scenario} & \textbf{Characteristics} & \textbf{Mass [kg]}\\\hline
1 & Baseline & 3.024\\
2 & Added masses: 75 g at 1010 mm, 12 g  at 1050 mm and 61 g at 1365 mm. & 3.172\\
3 & Added masses: 88 g at 1010 mm, 51 g  at 1050 mm, 83 g at 1205 mm and 61 g at 1365 mm. & 3.307\\
4 & Added masses: same as Scenario 2 plus 181 g at 570 mm and 170 g at 665 mm. & 3.658\\
\hline
\end{tabularx}
\end{table}

For each scenario, the wing was excited with a linear sine sweep excitation between 2 and 25 Hz lasting 20 min (single sweep - 0.02 Hzs\textsuperscript{-1} chirp rate) on a shaker table. The data was collected from a grid of 8 accelerometers placed across the wing span. The experimental data is available in \cite{Dessena2022g}.

The reader interested in more information on the XB-2 wing is referred to the following works: the second author's thesis \cite{Pontillo2020}, outlining its design and testing in the BeaRDS project framework, a GVT campaign in \cite{Dessena2022b}, a further GVT campaign with a revolving clamp \cite{Dessena2022e} a damping estimation from wind tunnel data in \cite{Tsatsas2022}, the development of an updated FEM in \cite{Dessena2024a}, and nonlinear dynamics identification in \cite{Dessena2022h}.

\subsection{Fast Relaxed Vector Fitting}\label{sec:FRVF}
The FRVF, first documented in its current and complete form in \citep{Knockaert2009}, originated as an improved version of the standard Vector Fitting (VF) algorithm, proposed by Gustavsen and Semlyen in 1999 for the modelling of large multiport electrical circuits \citep{Gustavsen1999a}. With respect to the VF algorithm, the FRVF adds a relaxed non-trivial constraint in the pole identification step \citep{Vector2006,Grivet2016} and exploits the matrix form of the linear problem and the QR decomposition, such that $\mathbf{A} = \mathbf{QR}$, for fast computation. In the form considered here, the FRVF procedure was tested and applied for the first time for the frequency domain SI of simple mechanical systems in \citep{Civera2021} and to large civil structures and infrastructures in \citep{Civera2021a}.
 
The Fast Relaxed Vector Fitting (FRVF) method presented here serves as an input-output system identification approach tailored for experimental modal analysis. It is versatile, accommodating any number of input and/or output channels and applicable to {different} input types \cite{Schwarz1999}{, such as random, periodic, and impulse.} While its original formulation (and the one adopted in this study) is expressed in the frequency domain, a time-domain variant is also reported in the literature \cite{Grivet-Talocia2003,Grivet-Talocia2004}. 

The FRVF algorithm approximates the Transfer Function (TF) between a measured output and a known input. This process accommodates any form of input (e.g., force, displacement, velocity, or acceleration) and output; however, for clarity, the relationship between the input force, $F$, and the output displacement, $Y$, is considered here. This specific relationship corresponds to the system receptance Frequency Response Function (FRF). To proceed, it is essential to revisit fundamental definitions. A pole of a mechanical system with a single degree of freedom can be described in the Laplace domain, i.e., within the complex $s$-plane, as:

\begin{equation}
s_{1,2} = \sigma \pm i\omega_d
\end{equation}

\noindent where, being a complex pair, $s_2 = s_1^*$. The receptance FRF can be approximated by a partial fraction form defined as:

\begin{equation}
H(s) = \frac{Y(s)}{F(s)} \approx \frac{A}{s - s_1} + \frac{A^*}{s - s_2}
\label{eq:2}
\end{equation}

\noindent {where A and its complex conjugate $A^*$ represent the residues of $H(s)$:}

\begin{equation}
A = +\frac{1}{i 2 m \omega_d}; A^* = -\frac{1}{i 2 m \omega_d};
\label{eq:A}
\end{equation}

{As the minus sign in the complex conjugate residue arises naturally from the conjugation ($i$ becoming $-i$). Notably, if one rationalises the imaginary unit in the denominator, multiplying by $i/i$, the first half of }\cref{eq:A}{ becomes $A = -\frac{i}{2 m \omega_d}$
which is a negative imaginary value; that is an important characteristic of the residue of a driving point.}

To express this TF in the frequency domain, note that the real part of $s_{1,2}$ can be represented as:

\begin{equation}
\sigma = - \zeta \omega_n
\end{equation}

\noindent while the imaginary component originates from the damped natural angular frequency $\omega_d$ of the corresponding mode:

\begin{equation}
\omega_d = \omega_n \sqrt{1 - \zeta^2}
\end{equation}

\noindent where $\zeta$ is the damping ratio and:

\begin{equation}
\omega_n = \sqrt{\frac{k}{m}}
\end{equation}

\noindent is the natural frequency of the $n$-th mode.

By evaluating $H(s)$ for $\sigma = 0$ (i.e. for a purely imaginary argument), the Laplace Transform is reduced to a Fourier Transform, and it can be shown that \cite{Mendes2009}:

\begin{equation}
H(\omega) = \left[ \frac{A}{i(\omega - \omega_d) + \zeta \omega_n} + \frac{A^*}{i(\omega + \omega_d) + \zeta \omega_n} \right] = \frac{1}{- m \omega^2 + i c \omega + k}
\end{equation}

It is then possible to determine the physical properties of the system — such as its mass $m$, stiffness $k$, and viscous damping $\bar{c} = 2m\zeta\omega_n$ — based on its vibrational response. By monitoring these properties over time while maintaining a constant mass, frequency shifts can be attributed to variations in stiffness and, consequently, the onset of damage.

The right-hand side of \cref{eq:2} can be generalised to multi-degree-of-freedom (MDoF) systems as a summation of partial fractions, expressed as:

\begin{equation}
f(s) = \sum_{n=1}^{N} \frac{A_n}{s - s_n} + H_{\infty}
\label{eq:fs=}
\end{equation}

\noindent where {$N$ is the number of modes assumed in the inspected range of the MDoF system (assumed since this is, of course, not known a prior) and} $H_{\infty}$ governs the behaviour at high frequencies and can be decomposed into two components: the real quantity $d$ and the term $se$ \cite{Gustavsen1999a}.{Concerning complex conjugate poles, VF supports both real and complex conjugate poles, thus not distinguishing automatically between vibrational (underdamped) modes, critically or overcritically (i.e. purely exponential) modes, and rigid body motions (real-valued poles at or near $s=0$). The removal of non-oscillatory modes, if needed, must be performed in post-processing after identification.} {Then, from }\cref{eq:fs=} an unknown function $\sigma(s)$ is defined as:

\begin{equation}
\sigma(s) \approx \sum_{n=1}^{N} \frac{\tilde{A}_n}{s - \tilde{s}_n} + 1
\end{equation}

\noindent Multiplying $f(s)$ by $\sigma(s)$ results in the following expression:

\begin{equation}
\sigma(s) f(s) \approx \left[ \sum_{n=1}^{N} \frac{A_n}{s - \tilde{s}_n} + d + se \right] \left[ \sum_{n=1}^{N} \frac{\tilde{A}_n}{s - \tilde{s}_n} + 1 \right]
\label{eq:sfs}
\end{equation}

\noindent The poles of $\sigma(s)f(s)$ are the same rational approximation of $\sigma(s)$. By forcing $\sigma(s)$ to approach unity at very high frequencies (that is to say, for $H_{\infty} = 1$), the ambiguity is removed from the solution for $\sigma(s)$ \cite{Gustavsen1999a}. {The terms $d$ and $se$ are heuristic approximations introduced to account for the contribution of dynamics outside the frequency band of interest -- similar in spirit to the upper and lower residuals in modal theory. However, they serve a more general-purpose modelling role.
The constant-value $d$ term (constant) captures low-frequency asymptotic behavior -- similar to a lower residual term $LR/s^2$. As $s\rightarrow0$, $H(s)\rightarrow d$, so it acts like a DC offset or residual representing static-like (quasi-static) gains that persists after dynamic modes are truncated. Conversely, the $se$ term (linearly proportional to frequency) dominates as $s\rightarrow +\infty$, capturing the high-frequency asymptotic behaviour. Hence, it is the counterpart to the upper residual term, accounting for the truncated higher-frequency modes not included in the modal summation. So, while both approaches serve a similar purpose -- modelling contributions from modes outside the inspected frequency band -- the modal residuals have a physical derivation, while the $d+se$ can be seen as empirical corrections in the rational function approximation framework. This lack of theoretical derivation makes them easily adaptable over different fields, from Electrical to Mechanical Engineering, while preserving their intended scope.} {Thus, }\cref{eq:sfs} leads to:

\begin{equation}
\sigma(s)f(s) \approx \left( \sum_{n=1}^{N} \frac{\tilde{A}_n}{s - \tilde{s}_n} + 1 \right) f(s) \approx \sum_{n=1}^{N} \frac{A_n}{s - \tilde{s}_n} + d + se
\end{equation}

\noindent where

\begin{equation}
(\sigma f)_{fit}(s) \approx (\sigma)_{fit}(s)f(s)
\end{equation}

\noindent After performing the necessary mathematical derivations, it can be shown that:

\begin{equation}
d + \sum_{n} \frac{A_n}{s - s_n} \approx H(s) \left( d' + \sum_{n} \frac{A_n'}{s - s_n} \right)
\end{equation}

This outlines the relationship between the TF $H(s)$ and the fitted residues $[A_n, A_n']$, poles $[s_n, s_n']$, and constant terms $[d, d']$. The right-hand side of Eq. (11) asymptotically converges to $H(s)$ as the solution is refined, which can be achieved iteratively by treating it as a least-squares (LS) optimisation problem.

This completes the fundamental definition of the VF method. Building upon this formulation, a Fast Relaxed VF (FRVF) implementation has been employed \cite{Gustavsen2006a}. This approach specifically introduces a relaxed non-triviality constraint in the pole identification process, incorporating a real-valued free variable $\hat{d}$:

\begin{equation}
\sigma(s) = \sum_{n=1}^{N} \frac{\tilde{A}_n}{s - s_n} + \hat{d}
\end{equation}

The only constraint is that the sum of the real parts of $\sigma(s)$ must remain non-zero to prevent a trivial solution. This straightforward modification provides two key benefits: it reduces the sensitivity to the initial pole placement by enabling larger shifts across the $s$-plane during the initial iterations and enhances the overall convergence of the solution.

The linear problem is subsequently addressed using the QR decomposition method, as proposed in \cite{Knockaert2009}. This involves decomposing the LS matrix into the product of an orthonormal basis matrix, $Q$, and an upper triangular matrix of coefficients, $R$. This approach substantially decreases both computational effort and memory storage requirements \cite{Grivet2016}. 
{In this work, we followed the initialisation process suggested in }\cite{Gustavsen1999a}{ (see subsection 3.2), for all the reasons explained in detail in the same text (subsections 4.2, 4.3, and 4.7). In brief, it is suggested there, and implemented here, to use complex conjugate pairs of initial poles, $-\alpha_n \pm i\beta_n$, with their imaginary parts $\beta$ evenly (linearly or logarithmically) spaced across the frequency range of interest. In this work, these were linearly spaced. Regarding the real part, this is initially assumed to be always negative (i.e., stable), assuming the initial values as $\alpha_n=\beta_n/100$, to ensure that the real parts are small enough to avoid numerical issues (as if initialised with overly fast decay, they can get numerically insignificant). 

For what concerns the iterative process, it is important to note that, in the classic VF, there is no stopping criterion. The convergence of the fitted function over experimental data needs to be manually checked by the user. Furthermore, this does not enforce that the identified poles are stable. Theoretically, any negative, positive, and null real parts can be identified. Hence, once again, checks, this time for stability, are left to the user.
The stability and system order problems can be addressed, as commonly done in practical modal analysis application, with the implementation of stabilisation diagrams, which minimum order is equal to the expected -- by experience -- minimum number of modes and the maximum order by a number of states which allows for the identification of all modes of interest.}
Further details regarding the implementation and the associated technical considerations can be found in \cite{Gustavsen1999a,Grivet2016}, for the VF and FRVF, respectively.

\subsection{Loewner Framework}\label{sec:LF}
Previously, the LF algorithm has been employed for the modelling of multi-port electrical systems \cite{Lefteriu2009} and utilised for aerodynamic model order reduction in the context of aeroservoelastic modelling \cite{Quero2019}. Later, the first and second authors applied the LF for the identification of modal parameters from SIMO mechanical systems in \cite{Dessena2022}, verified its computational efficiency in \cite{Dessena2022f}, and assessed its robustness to noise for SHM in \cite{Dessena2022g}. Further developments have included the extension of the LF for the extraction of modal parameters from multi-input multi-output \cite{Dessena2024,Dessena2024c} and output-only systems \cite{Dessena2024f}. Nevertheless, the version considered in this work is the SIMO version first introduced in \cite{Dessena2022}.

In order to properly introduce the LF, let us begin by defining the Loewner matrix ${\bm{\mathbb{L}}}$:
\noindent \emph{Given a row array of pairs of complex numbers ($\mu_j$,${v}_j$), $j=1$,...,$q$, and a column array of pairs of complex numbers ($\lambda_i$,${w}_j$), $i=1$,...,$k$, with $\lambda_i$, $\mu_j$ distinct, the associated $\boldsymbol{\bm{\mathbb{L}}}$, or divided-differences matrix is:}
\begin{equation}
\label{eq:LM}
\boldsymbol{\bm{\mathbb{L}}}=\begin{bmatrix}
\frac{\bm{v}_1-\bm{w}_1}{\mu_1-\lambda_1} & \cdots & \frac{\bm{v}_1-\bm{w}_k}{\mu_1-\lambda_k}\\
\vdots & \ddots & \vdots\\
\frac{\bm{v}_q-\bm{w}_1}{\mu_q-\lambda_1} & \cdots & \frac{\bm{v}_q-\bm{w}_k}{\mu_q-\lambda_k}\\
\end{bmatrix}\:\in \mathbb{C}^{q\times k}
\end{equation}
\emph{If there is a known underlying function $\pmb{\phi}$, then $\bm{w}_i=\pmb{\phi}(\lambda_i)$ and $\bm{v}_j=\pmb{\phi}(\mu_j).$}

Karl Löwner established a relationship between $\boldsymbol{\mathbb{L}}$ and rational interpolation, often referred to as Cauchy interpolation \cite{Lowner1934}. This connection allows interpolants to be defined through the determinants of submatrices of $\boldsymbol{\mathbb{L}}$. As demonstrated in \cite{Antoulas2017, Mayo2007}, rational interpolants can be directly derived from $\boldsymbol{\mathbb{L}}$. This study adopts the methodology based on the Loewner pencil, which consists of the matrices $\boldsymbol{\mathbb{L}}$ and $\boldsymbol{\mathbb{L}}_s$. Here, $\boldsymbol{\mathbb{L}}_s$ denotes the \emph{Shifted Loewner matrix}, which will be introduced later.

To illustrate the working principle of the LF, consider a linear time-invariant dynamical system $\bm{\Sigma}$ characterised by $m$ inputs, $p$ outputs, and $k$ internal variables, expressed in descriptor form as:

\begin{equation}
\bm{\Sigma}:\;\bm{E}\frac{d}{dt}\bm{x}(t)=\bm{A}\bm{x}(t)+\bm{B}\bm{u}(t);\;\;\;
\bm{y}(t)=\bm{C}\bm{x}(t)+\bm{D}\bm{u}(t)
 \label{eq:LTI}
 \end{equation}
\noindent where $\bm{x}(t) \in \mathbb{R}^{k}$ represents the vector of internal variables, $\bm{u}(t) \in \mathbb{R}^{m}$ is the input function, and $\bm{y}(t) \in \mathbb{R}^{p}$ denotes the output vector. The following constant matrices characterise the system:

\begin{equation}
    \bm{E},\bm{A}\in \mathbb{R}^{k\times k},\; \bm{B}\in \mathbb{R}^{k\times m}\; \bm{C}\in \mathbb{R}^{p\times k}\; \bm{D}\in \mathbb{R}^{p\times m}
\end{equation}
\noindent The Laplace transfer function, $\bm{H}(s)$, associated with $\bm{\Sigma}$ can be formulated as a $p \times m$ rational matrix function, under the condition that the matrix $\bm{A} - \lambda\bm{E}$ remains non-singular for a given finite value of $\lambda$, where $\lambda \in \mathbb{C}$:

\begin{equation}
    \bm{H}(s)=\bm{C}(s\bm{E}-\bm{A})^{-1}\bm{B}+\bm{D}
    \label{eq:trans}
\end{equation}
\noindent Let us consider the general framework of tangential interpolation, often referred to as rational interpolation in tangential directions \cite{Kramer2016}. The corresponding right interpolation data is defined as:

\begin{equation}
\begin{gathered}
   (\lambda_i;\bm{r}_i,\bm{w}_i),\: i = 1,\dots,\rho
    \\
    \begin{matrix}
        \bm{\Lambda}=\text{diag}[\lambda_1,\dotsc,\lambda_k]\in \mathbb{C}^{\rho\times \rho}\\
        \bm{R}=[\bm{r}_1\;\dotsc \bm{r}_k]\in \mathbb{C}^{m\times \rho}\\
        \bm{W} = [\bm{w}_1\;\dotsc\;\bm{w}_k]\in \mathbb{C}^{\rho\times \rho}
        \end{matrix}\Bigg\}
\end{gathered}
\label{eq:RID}
\end{equation}
\noindent In a similar manner, the left interpolation data is defined.
\begin{equation}
    \begin{gathered} 
        (\mu_j,\bm{l}_j,\bm{v}_j),\: j = 1,\dots,v
        \\ 
        \begin{matrix}
            \bm{M}=\mathrm{diag}[\mu_1,\dotsc,\mu_q]\in \mathbb{C}^{v\times v}\\
            \bm{L}^T=[\bm{l}_1\;\dotsc \bm{l}_v]\in \mathbb{C}^{p\times v}\\
            \bm{V}^T = [\bm{v}_1\;\dotsc\;\bm{v}_q]\in \mathbb{C}^{m\times v}
        \end{matrix}\Bigg\}
    \end{gathered}
\label{eq:LID}
\end{equation}
\noindent The values $\lambda_i$ and $\mu_j$ correspond to the points where $\bm{H}(s)$ is evaluated, representing the frequency bins in this context. The vectors $\bm{r}_i$ and $\bm{l}_j$ define the right and left tangential directions, which are commonly selected randomly in practice \cite{Quero2019}, while $\bm{w}_i$ and $\bm{v}_j$ represent the associated tangential data. The rational interpolation problem is resolved by establishing a link between $\bm{w}_i$ and $\bm{v}_j$ and the transfer function $\bm{H}$, associated with the realisation $\bm{\Sigma}$ in \cref{eq:LTI}:

\begin{equation}
\bm{H}(\lambda_i)\bm{r}_i=\bm{w}_i,\:j=1.\dots,\rho \;\; \text{and} \;\;
\bm{l}_i\bm{H}(\mu_j)=\bm{v}_j,\:i=1,\dots,v
 \label{eq:LS2}
 \end{equation}
\noindent ensuring that the Loewner pencil satisfies \cref{eq:LS2}. 
Now, let us consider a set of points $Z = \{z_1, \dots, z_N\}$ in the complex plane and a rational function $\bm{y}(s)$, where $\bm{y}_i := \bm{y}(z_i)$ for $i = 1, \dots, N$, with $Y = \{\bm{y}_1, \dots, \bm{y}_N\}$. Incorporating the left and right data partitions yields the following expressions:

\begin{equation}
\begin{split}
	Z=\{\lambda_1,\dots,\lambda_\rho\} \cup \{\mu_1,\dots,\mu_v\} \text{and} 
	Y=\{\bm{w}_1,\dots,\bm{w}_\rho\} \cup \{\bm{v}_1,\dots,\bm{v}_v\}
\end{split}
\label{eq:ZY}
\end{equation}
\noindent where $N = p + v$. As a result, the matrix $\boldsymbol{\mathbb{L}}$ is expressed as:
\begin{equation}
\label{eq:LM2}
\boldsymbol{\bm{\mathbb{L}}}=\begin{bmatrix}
\frac{\bm{v}_1\bm{r}_1-\bm{l}_1\bm{w}_1}{\mu_1-\lambda_1} & \cdots & \frac{\bm{v}_1\bm{r}\rho-\bm{l}_1\bm{w}\rho}{\mu_1-\lambda\rho}\\
\vdots & \ddots & \vdots\\
\frac{\bm{v}_v\bm{r}_1-\bm{l}_v\bm{w}_1}{\mu_v-\lambda_1}& \cdots & \frac{\bm{v}_v\bm{r}\rho-\bm{l}_v\bm{w}\rho}{\mu_v-\lambda\rho}\\
\end{bmatrix}\:\in \mathbb{C}^{v\times \rho}
\end{equation}

\noindent Since $\bm{v}_v\bm{r}_p$ and $\bm{l}_v\bm{w}_p$ are scalars, the Sylvester equation for $\boldsymbol{\mathbb{L}}$ is satisfied as follows:
\begin{equation}
    \boldsymbol{\bm{\mathbb{L}}}\bm{\Lambda}-\bm{M}\boldsymbol{\bm{\mathbb{L}}}=\bm{L}\bm{W}-\bm{V}\bm{R}
    \label{eq:syl}
\end{equation}

\noindent The \emph{shifted Loewner matrix}, $\boldsymbol{\mathbb{L}}_s$, is defined as the matrix $\boldsymbol{\mathbb{L}}$ associated with $s\bm{H}(s)$:

\begin{equation}
\label{eq:LS}
\boldsymbol{\bm{\mathbb{L}}}_s=\begin{bmatrix}
\frac{\mu_1\bm{v}_1\bm{r}_1-\lambda_1\bm{l}_1\bm{w}_1}{\mu_1-\lambda_1} & \cdots & \frac{\mu_1\bm{v}_1\bm{r}_\rho-\lambda_\rho\bm{l}_1\bm{w}_\rho}{\mu_1-\lambda_\rho}\\
\vdots & \ddots & \vdots\\
\frac{\mu_v\bm{v}_v\bm{r}_1-\lambda_1\bm{l}_v\bm{w}_1}{\mu_v-\lambda_1}& \cdots & \frac{\bm{v}_v\bm{r}_\rho-\bm{l}_v\bm{w}_\rho}{\mu_v-\lambda_\rho}\\
\end{bmatrix}\:\in \mathbb{C}^{v\times \rho}
\end{equation}

\noindent Similarly, the Sylvester equation is satisfied as follows:
\begin{equation}
    \boldsymbol{\bm{\mathbb{L}}}_s\Lambda-\bm{M}\boldsymbol{\bm{\mathbb{L}}}_s=\bm{L}\bm{W}\bm{\Lambda}-\bm{M}\bm{V}\bm{R}
    \label{eq:syl2}
\end{equation}

Without loss of generality, $\bm{D}$ can be assumed to be zero, as its contribution does not affect the tangential interpolation within the LF framework \cite{Mayo2007}. For simplicity, the following discussion will focus on $\bm{H}(s)$. As a result, \cref{eq:trans} reduces to:

\begin{equation}
\bm{H}(s)=\bm{C}(s\bm{E}-\bm{A})^{-1}\bm{B}
\label{eq:fin}
\end{equation}

A minimal-dimensional realisation can be achieved only if the system is fully controllable and observable. Assuming that the data is sampled from a system whose transfer function is described by \cref{eq:fin}, the generalised tangential observability, $\mathcal{O}_v$, and generalised tangential controllability, $\mathcal{R}_\rho$, are defined in \cite{Lefteriu2010b}. Consequently, \cref{eq:LM2} and \cref{eq:LS} can be reformulated as:

\begin{align}
    \boldsymbol{\bm{\mathbb{L}}}=-\mathcal{O}_v\bm{E}\mathcal{R}_\rho &&
    \boldsymbol{\bm{\mathbb{L}}}_s=-\mathcal{O}_v\bm{A}\mathcal{R}_\rho
    \label{eq:LLs}
\end{align}
Then, by defining the Loewner pencil as a regular pencil, such that $\mathrm{eig}((\boldsymbol{\bm{\mathbb{L}}},\boldsymbol{\bm{\mathbb{L}}}_s)) \neq (\mu_i, \lambda_i)$:
\begin{align}
    \bm{E}=-\boldsymbol{\bm{\mathbb{L}}},&&\bm{A}=-\boldsymbol{\bm{\mathbb{L}}}_s,&&\bm{B}=\bm{V},&&\bm{C}=\bm{W}
\end{align}
Consequently, the interpolating rational function can be expressed as:
\begin{equation}
    \bm{H}(s)=\bm{W}(\boldsymbol{\bm{\mathbb{L}}}_s-s\boldsymbol{\bm{\mathbb{L}}})^{-1}\bm{V}
\end{equation}
The derivation presented applies specifically to the minimal data scenario, which is seldom encountered in practical applications. Nevertheless, the LF framework can be extended to accommodate redundant data points efficiently. To proceed, consider the assumption:
\begin{equation}
\begin{split}
    \mathrm{rank}[\zeta\boldsymbol{\bm{\mathbb{L}}}-\boldsymbol{\bm{\mathbb{L}}}_s]=\mathrm{rank}[\boldsymbol{\bm{\mathbb{L}}}\:\boldsymbol{\bm{\mathbb{L}}}_s]
   =\mathrm{rank}
    \begin{bmatrix}
    \boldsymbol{\bm{\mathbb{L}}} \boldsymbol{\bm{\mathbb{L}}}_s
    \end{bmatrix}=k,\; 
    \forall \zeta \in \{\lambda_j\}\cup\{\mu_i\}
    \end{split}
    \label{eq:cond1}
\end{equation}
Next, a short Singular Value Decomposition (SVD) is performed on $\zeta\boldsymbol{\mathbb{L}} - \boldsymbol{\mathbb{L}}_s$:
\begin{equation}
    \textrm{svd}(\zeta\boldsymbol{\bm{\mathbb{L}}}-\boldsymbol{\bm{\mathbb{L}}}_s)=\bm{Y}\bm{\Sigma}_l\bm{X}
\label{eq:cond2}
\end{equation}
where $\mathrm{rank}(\zeta\boldsymbol{\bm{\mathbb{L}}}-\boldsymbol{\bm{\mathbb{L}}}_s)=\mathrm{rank}(\bm{\Sigma}_l)=\mathrm{size}(\bm{\Sigma}_l)=k,\bm{Y}\in\mathbb{C}^{v \times k}$ and $\bm{X}\in\mathbb{C}^{k\times \rho}$.
Note that:
\begin{equation}
\begin{split}
-\bm{A}\bm{X}+\bm{E}\bm{X}\bm{\Sigma}_l = \bm{Y}^*\boldsymbol{\bm{\mathbb{L}}}_s\bm{X}^*\bm{X}-\bm{Y}^*\boldsymbol{\bm{\mathbb{L}}}\bm{X}^*\bm{X}\bm{\Sigma}_l=\bm{Y}^*(\boldsymbol{\bm{\mathbb{L}}}_s-\boldsymbol{\bm{\mathbb{L}}}\bm{\Sigma}_l )=\bm{Y}^*\bm{V}\bm{R}=\bm{BR}
\end{split}
\label{eq:cond3}
\end{equation}
In a similar fashion, the relationship $-\bm{Y}\bm{A} + \bm{M}\bm{Y}\bm{E} = \bm{L}\bm{C}$ holds, where $\bm{X}$ and $\bm{Y}$ represent the generalised controllability and observability matrices, respectively, for the system $\bm{\Sigma}$, assuming $\bm{D} = 0$.

After verifying that the right and left interpolation conditions are satisfied, the Loewner realisation incorporating redundant data can be formulated as:
\begin{equation}
\begin{split}
\bm{E}=-\bm{Y}^*\boldsymbol{\bm{\mathbb{L}}}\bm{X},\quad \quad
\bm{A}=-\bm{Y}^*\boldsymbol{\bm{\mathbb{L}}}_s\bm{X,} \quad \quad
\bm{B}=\bm{Y}^*\bm{V},\quad \quad
\bm{C}=\bm{W}\bm{X}
\end{split}
\label{eq:real}
\end{equation}

The formulation in \cref{eq:real}, which defines the Loewner realisation for redundant data, will serve as the foundation throughout this work. For a detailed explanation of each step, readers are referred to \cite{Mayo2007,Antoulas2017}, while the MATLAB implementation is available in \cite{Dessena2021}. Finally, the system modal parameters can be extracted through eigenanalysis of the system matrices $\bm{A}$ and $\bm{C}$ in \cref{eq:real}. {In its practical application to modal analysis, this is carried out for a series of model order $k$ such that to include the minimum number of modes of interest and the maximum order at which these modes are stably identified.}

\subsection{The Simplified Aeroelastic Model}\label{sec:aero}
Aeroelasticity is concerned with the interaction between aerodynamic, elastic, and inertia loads. This interaction can arise in an unstable manner, hence generating the so-called aeroelastic phenomena \citep{Li2023}. In general, aeroelasticity can be divided into two subgroups: static and dynamic \citep{Wright2014}. The former includes divergence and aileron reversal, while the latter mainly concerns flutter, sub-critical dynamics, and limit cycle oscillations. The division arises from the fact that static phenomena can depend solely on quasi-static behaviours, while flutter has a harmonic dependence. The reader interested in a more profound review on the subject is referred to \citep{Sudha2020} for a focus on flutter prediction techniques and to \citep{Amato2019} for a practical application. 

\begin{figure}[!ht]
\centering
    \includegraphics[width=.6\textwidth, keepaspectratio]{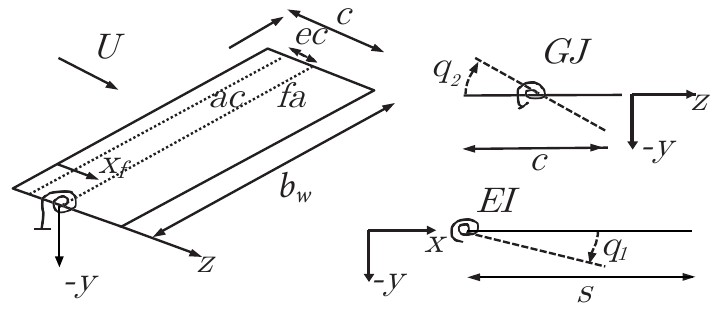}
    \caption{Flexible wing 2 DoF model schematic.\label{fig:wing}}
\end{figure}

In this work, an aeroelastic model is developed to assess, respectively, the static and dynamic aeroelasticity phenomena of different experimental configurations. The main focus is to detect divergence and flutter onset speeds. The proposed model is based on the first spanwise flapping (i.e., bending) and the first twisting (torsion) modes only. For simplicity, an aeroelastic model for a rectangular flexible wing is used; so, a 2 DoF model based on oscillatory aerodynamics, an extension for flexible wings of the classic binary aeroelastic model \citep{Wright2014}, is selected. {Nevertheless, this should not be seen as a method limitation, but rather as a modelling choice, since more system modes can be considered by considering more DoFs.}
{Before starting with the model description, the following assumptions need to be considered}
\begin{itemize}
    \item {The wing shape is assumed to be rectangular;}
    \item {The flexural stiffness EI, the torsional stiffness GJ, and the mass distribution are assumed to remain constant along the span;}
    \item {The identified $\zeta_n$ are assumed to be dependent only on structural effects in the wind-off results;}
    \item {XB-2 second mode (1\textsuperscript{st} coupled mode) is assumed to be a pure twisting mode.}
\end{itemize}
First, let us define the deflection, $z$, of a point ($x$, $y$) on the wing such as:

\begin{equation}
    z = y^2q_1+y(x-x_f)q_2
    \label{eq:z}
\end{equation}
where $q_1$ and $q_2$ are the generalized coordinates corresponding to the flapping deflection and the pitch angle.
Then, {by applying the Lagrangian method based on the estimation of kinetic and potential energy and incremental work}, the following equations are obtained \citep{Wright2014} from \cref{eq:z}:

\begin{multline}
    m
    \begin{bmatrix}
    \frac{\bar{c}b_w^5}{5} & \dfrac{b_w^4}{4}\left(\dfrac{\bar{c}^2}{2}-\bar{c}x_f \right) \\
    \dfrac{b_w^4}{4} \left( \dfrac{\bar{c}^2}{2} - \bar{c} x_f \right) & \dfrac{b_w^3}{3} \left(\dfrac{\bar{c}^3}{3} - \bar{c}^2 x_f + \bar{c} x^2_f \right)
    \end{bmatrix}
    \begin{Bmatrix}
    \ddot q_1 \\ \ddot q_2 
    \end{Bmatrix} + \\+\rho U_\infty
    \begin{bmatrix}
    \dfrac{\bar{c}a_wb_w^5}{10} & 0 \\
    -\dfrac{\bar{c}^2ea_wb_w^4}{8} & -\dfrac{\bar{c}^3b_w^3M_{\dot \theta}}{24}  
    \end{bmatrix}
    \begin{Bmatrix}
    \dot q_1 \\ \dot q_2 
    \end{Bmatrix}+\\
    +\left(\rho U_\infty
    \begin{bmatrix}
    0 & \dfrac{\bar{c}a_wb_w^4}{8}\\
    0 & -\dfrac{\bar{c}^2ea_wb_w^3}{6}
    \end{bmatrix}+
    \begin{bmatrix}
    4EIb_w & 0 \\
    0 & GJb_w
    \end{bmatrix}
    \right)
    \begin{Bmatrix}
    q_1 \\ q_2 
    \end{Bmatrix}=
     \begin{Bmatrix}
    0 \\ 0 
    \end{Bmatrix}
    \label{eq:ae_model}
\end{multline}

\noindent where $m$ is mass per unit area in kgm\textsuperscript{-2}, $\bar{c}$ is the wing chord {(as the XB-2 mean aerodynamic chord)} in m, $b_w$ is the wing span in m, $x_f$ is the flexural axis position with respect to $\bar{c}$, $\rho$ is the air density in kgm\textsuperscript{-3}, $U_\infty$ is the air speed in ms\textsuperscript{-1}, $a_w$ is the lift curve slope, $EI$ is the bending, or flapping, stiffness, $GJ$ is the torsional stiffness, and $M_{\dot \theta}$ is the non-dimensional pitch damping derivative{, which is estimated following Theodorsen’s unsteady
aerodynamics.} \Cref{fig:wing} shows the schematic for the wing ROM.

\noindent \Cref{eq:ae_model} is a special form, for the zero {viscous} damping case ${\mathbf{d}=0}$, of the following:
\begin{equation}
    \mathbf{a_m}  \ddot{\mathbf{q}} +\left(\rho V \mathbf{b} + \mathbf{d}\right)\dot{\mathbf{q}} +\left(\rho V^2 \mathbf{c}_a + {\mathbf{e}}\right){\mathbf{q}}=\mathbf{0}
    \label{eq:ful}
\end{equation}
where $\mathbf{a_m} $ is the mass, or inertia, matrix, $\mathbf{b}$ is the aerodynamic damping matrix, $\mathbf{c}_a$ is the aerodynamic stiffness matrix, $\mathbf{d}$ is the {viscous} damping and ${\mathbf{e}}$ is the stiffness matrix. Now, considering the availability of modal data{, it can be shown (see }\cite{Coleman1940}{) that for low-amplitudes GVTs of beam-like structures, like wings, the viscous damping is comparable to a constant-in-frequency hysteric damping (forming graphically a perfect ellipse), which yields similar results to a structural damping model }\cite{Soroka1949}{. Furthermore, the identified damping value is considered constant across the frequency of interest, as the modelling approach chosen assumes small amplitudes (damping stays independent from vibration amplitude) and that the change in the aerodynamic damping along the frequency is dominant compared to a potential change in the viscous damping. The latter is a standard assumption for wind-off GVT in aeroelasticity }\cite{Wright2014}. $\mathbf{d}$ can be {considered as viscous damping and }built from the uncoupled modal damping assumption in \cite{Naeim1999}:

\begin{equation}
    \mathbf{d} = \bm{\phi}^{-T}\mathbf{d}_n\bm{\phi}^{-1}
    \;\text{  for  }\;
    \mathbf{d}_n = 2\zeta_n\omega_n a_n
    \label{eq:D}
\end{equation}
where the subscript $n$ identifies the uncoupled matrix, $\omega_n$ the natural frequency, $\zeta_n$ the damping ratio and $\bm{\phi}$ the mode shape. Hence, \cref{eq:ae_model,eq:D} can be combined to assemble \cref{eq:ful}.
Given full knowledge of the wing geometric characteristics, three properties remain to be defined: $M_{\dot \theta}$, $EI$, and $GJ$. $M_{\dot \theta}$ is the unsteady aerodynamics term and it is defined from oscillatory aerodynamics \citep{Wright2014}:

\begin{equation}
    M_{\dot{\theta}} = 2\pi\Bigg[ -\dfrac{f_k}{2}\left(\dfrac{1}{2}-a\right)+f_kF\left(a+\dfrac{1}{2}\right)\left(\dfrac{1}{2}-a\right)
    +\dfrac{G}{f_k}\Bigg(\dfrac{1}{2}+a\Bigg) \Bigg]
\label{eq:Mtheta}
\end{equation}
where $f_k$ is the reduced frequency, $a$ is the ratio between $\bar{c}$ and the flexural axis position, and $F$ and $G$ are, respectively, the real and imaginary part of Theodorsen's function, $C(f_k)$, such that:
\begin{equation}
    C(f_k)=F(f_k)+jG(f_k)=\dfrac{H^{(2)}_1(f_k)}{H^{(2)}_1(f_k)+jH^{(2)}_1(f_k)}
\end{equation}
where $H^{(2)}_n(f_k)$ are Hankel functions of the second kind and {$j$ is the imaginary number}. Note that in \cref{eq:Mtheta}, $2\pi$ represents the lift curve slope $a_w$. In this work, the actual slope value is considered rather than its approximation.

Concerning the bending and torsional stiffness, $EI$ and $GJ$, let us consider the still air case where $\mathbf{b}$ and $\mathbf{c}_a$ are zero. \cref{eq:ful} then becomes a simple mass-spring-damper system:

\begin{equation}
    \mathbf{a_m}  \ddot{\mathbf{q}} +\mathbf{d}\dot{\mathbf{q}} +\mathbf{e}\dot{\mathbf{q}}=\mathbf{0}
    \label{eq:wo}
\end{equation}

The natural frequencies can then be easily extracted through eigenanalysis. Hence, by having a set of experimental $\omega_n$ it is possible to define the $EI$ and $GJ$ of the equivalent system by minimising its squared difference to the experimental $\omega_n$. {Thus, by using this and the identified $\zeta_n$ for }\cref{eq:D}, an aeroelastic model can be defined starting from experimental data.

To study the system stability, the eigenanalysis of \cref{eq:wo} can be solved iteratively with the well-known $p$-$k$ method \citep{Gu2012} to find the divergence and flutter onset speeds.
The $p$-$k$ method is based on the hypothesis that pure harmonic aerodynamics can be used as a good approximation for lightly damped harmonic motions. This allows the computation of the aerodynamic transfer matrix at a complex frequency $p = \delta \pm jk$, such that $p \approx jk$. In simple terms, the real part, i.e. the damping, is neglected. A widely accepted workflow of the $p$-$k$ method~\citep{Wright2014} can be summarised as follow:

\begin{enumerate}
    \item Initiate an estimation, usually the still air value, of $p$, said $p_0 = \delta \pm jk_0$
    \item Evaluate the aerodynamics, in our case $M_{\dot \theta}$
    \item Solve the eigenvalue ($\lambda$) problem for \cref{eq:ful} and obtain a new set of $\lambda$, $p_1 = \delta \pm jk_1$
    \item Iterate between 2 and 3 until $k_n \approx k_{n-1}$
\end{enumerate}

From the $\lambda$ obtained after convergence, it is possible to build $\omega_n$, $\zeta_n$, real($\lambda$) and imag($\lambda$) vs air-speed ($U_\infty$) plots, which can be used to graphically portray divergence or flutter speed, whichever is detected first. Particularly, critical speeds are identified for $\zeta_n$ approaching zero or for real($\lambda$) zero crossings, since both cases are interpreted as instability in the system. Particularly, for flutter, only the real($\lambda$) zero crossing condition needs to be satisfied, while for divergence, $\text{imag}(\lambda)$ must be zero.

\section{Results\label{sec:res}}
The flutter speed prediction is carried out using the model introduced in \cref{sec:aero}, which is built on the modal parameters {($\omega_n$ and $\zeta_n$)} identified in \cite{Dessena2022g} via N4SID, FRVF, and LF and reported here in \cref{tab:freq}. 

\begin{table*}[!hbt]
\centering
\caption{Natural frequencies and damping ratios identified by N4SID, LF, and FRVF for all scenarios with relative differences (in parentheses). Data retrieved from \cite{Dessena2022g}.\label{tab:freq}}
\resizebox{\textwidth}{!}{ 
\begin{tabular}{lccccccccc}
\toprule
\multicolumn{10}{c}{\textbf{Natural Frequency} [Hz] (Difference wrt N4SID [\%])}\\\hline
\textbf{Mode}	& \multicolumn{3}{c}{1\textsuperscript{st} Bending}	& \multicolumn{3}{c}{1\textsuperscript{st} Coupled} & \multicolumn{3}{c}{2\textsuperscript{nd} Coupled}\\\midrule
\textbf{Scenario} & N4SID & LF & FRVF & N4SID & LF & FRVF & N4SID & LF & FRVF \\\hline
\textbf{1} & 3.190 & 3.202 & 3.203 & 11.896 & 11.886 & 11.858 & 17.763 & 17.703 & 17.725 \\
           & --    & (0.38) & (0.41) & --    & (-0.08) & (-0.32) & --    & (-0.34) & (-0.21) \\\hline
\textbf{2} & 2.957 & 2.958 & 2.945 & 12.096 & 12.134 & 12.083 & 17.350 & 17.302 & 17.294 \\
           & --    & (0.03) & (-0.41) & --    & (0.31) & (-0.11) & --    & (-0.28) & (-0.32) \\\hline
\textbf{3} & 2.775 & 2.769 & 2.788 & 12.002 & 12.025 & 12.014 & 17.079 & 17.101 & 17.023 \\
           & --    & (-0.22) & (0.47) & --    & (0.19) & (0.10) & --    & (0.13) & (-0.33) \\\hline
\textbf{4} & 2.729 & 2.725 & 2.727 & 11.970 & 11.965 & 11.938 & 15.067 & 15.052 & 15.004 \\
           & --    & (-0.15) & (-0.07) & --    & (-0.04) & (-0.27) & --    & (-0.10) & (-0.42) \\\hline
\multicolumn{10}{c}{\textbf{Damping Ratio} [-] (Difference wrt N4SID [\%])}\\\hline
\textbf{Mode}	& \multicolumn{3}{c}{1\textsuperscript{st} Bending}	& \multicolumn{3}{c}{1\textsuperscript{st} Coupled} & \multicolumn{3}{c}{2\textsuperscript{nd} Coupled}\\\midrule
\textbf{Scenario} & N4SID & LF & FRVF & N4SID & LF & FRVF & N4SID & LF & FRVF \\\hline
\textbf{1} & 0.032 & 0.040 & 0.028 & 0.066 & 0.063 & 0.065 & 0.058 & 0.061 & 0.062 \\
           & --    & (25.00) & (-12.50) & --    & (-4.55) & (-1.52) & --    & (5.17) & (6.90) \\\hline
\textbf{2} & 0.021 & 0.024 & 0.025 & 0.060 & 0.057 & 0.058 & 0.061 & 0.056 & 0.060 \\
           & --    & (14.29) & (19.05) & --    & (-5.00) & (-3.33) & --    & (-8.20) & (-1.64) \\\hline
\textbf{3} & 0.019 & 0.022 & 0.021 & 0.058 & 0.055 & 0.057 & 0.050 & 0.050 & 0.057 \\
           & --    & (15.79) & (10.53) & --    & (-5.17) & (-1.72) & --    & (0.00) & (14.00) \\\hline
\textbf{4} & 0.019 & 0.021 & 0.019 & 0.050 & 0.048 & 0.052 & 0.046 & 0.039 & 0.038 \\
           & --    & (10.53) & (0.00) & --    & (-4.00) & (4.00) & --    & (-15.22) & (-17.39) \\\hline
\end{tabular}}
\end{table*}

The comparison assumes the N4SID parameters as the benchmark values to assess the LF and FRVF performance. 

To accommodate the model assumptions (see \cref{sec:aero}), the mean aerodynamic chord is taken under consideration and, since the wing is assumed to be rectangular; the stiffnesses, EI and GJ, are also assumed to be constant along the span. The $\zeta_n$ assumption stands as a good approximation for small vibrations \citep{Civera2021b}. Lastly, the proposed model utilises the first bending mode exclusively for the bending component. Meanwhile, the first mode that exhibits torsional motion, even if coupled, is treated as pure torsion for the purposes of this analysis. The assumptions are motivated by the fact that the main goal of the aeroelastic investigation is to compare the LF and FRVF, rather than developing a full aeroelastic assessment. Nevertheless, similar approaches still give reasonable estimates, usually underestimating flutter onset speed by around 20\% \citep{keane2017}. Thus, according to the wing geometry and physical properties, \cref{eq:ae_model,eq:ful} are populated with the values in \cref{tab:ae_data}{, resulting in the first reduced order (2-DoF) dynamical model developed for XB-2.}

\begin{table}[!hbt]
\caption{\label{tab:ae_data} XB-2: Properties values for the aeroelastic model.}
\centering
\begin{tabular}{ll}
\hline
\textbf{Property} & \textbf{Value}\\\hline
m & mass divided by area \\ 
$\rho$ & 1.225 kgm\textsuperscript{-3} \\ 
$s_f$ & 0.25 $\times$ c\\
$\bar{c}$ & 172 mm \\ 
$a_w$ & 7.143 (NACA 23015) \\ 
$e$ & 0 \\
$b_w$ & 1.385 m \\
\hline
\end{tabular}
\end{table}

The aeroelastic system is then evaluated, using the $p-k$ method, between 0 and 28 ms\textsuperscript{-1} $U_\infty$, EI and GJ are derived from still air results (the GVT results) and $M_{\dot \theta}$ is a function of the reduced frequency $k${, being the first time that FRVF and LF-identified modal parameters are used to inform a flutter estimation process.} 
\begin{figure}[!ht]
\centering
    \begin{subfigure}[b]{.49\textwidth}
		\centering
      \includegraphics[width=.98\textwidth]{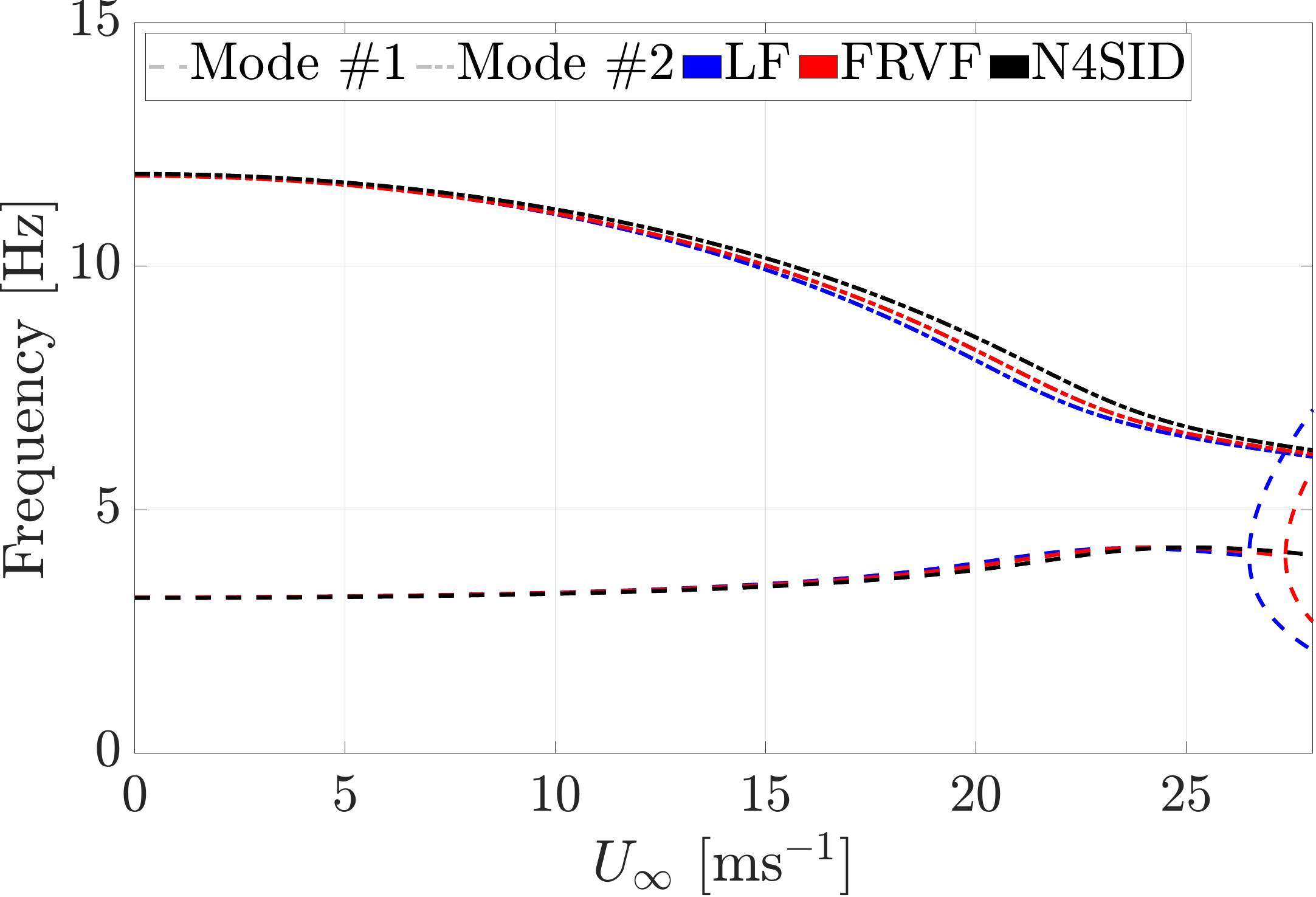}
      	\captionsetup{font={it},justification=centering}
		\subcaption{}
		\label{fig:flut1a}	
	\end{subfigure}
   \vspace{1em}
	\begin{subfigure}{.49\textwidth}
		\centering
       \includegraphics[width=\textwidth]{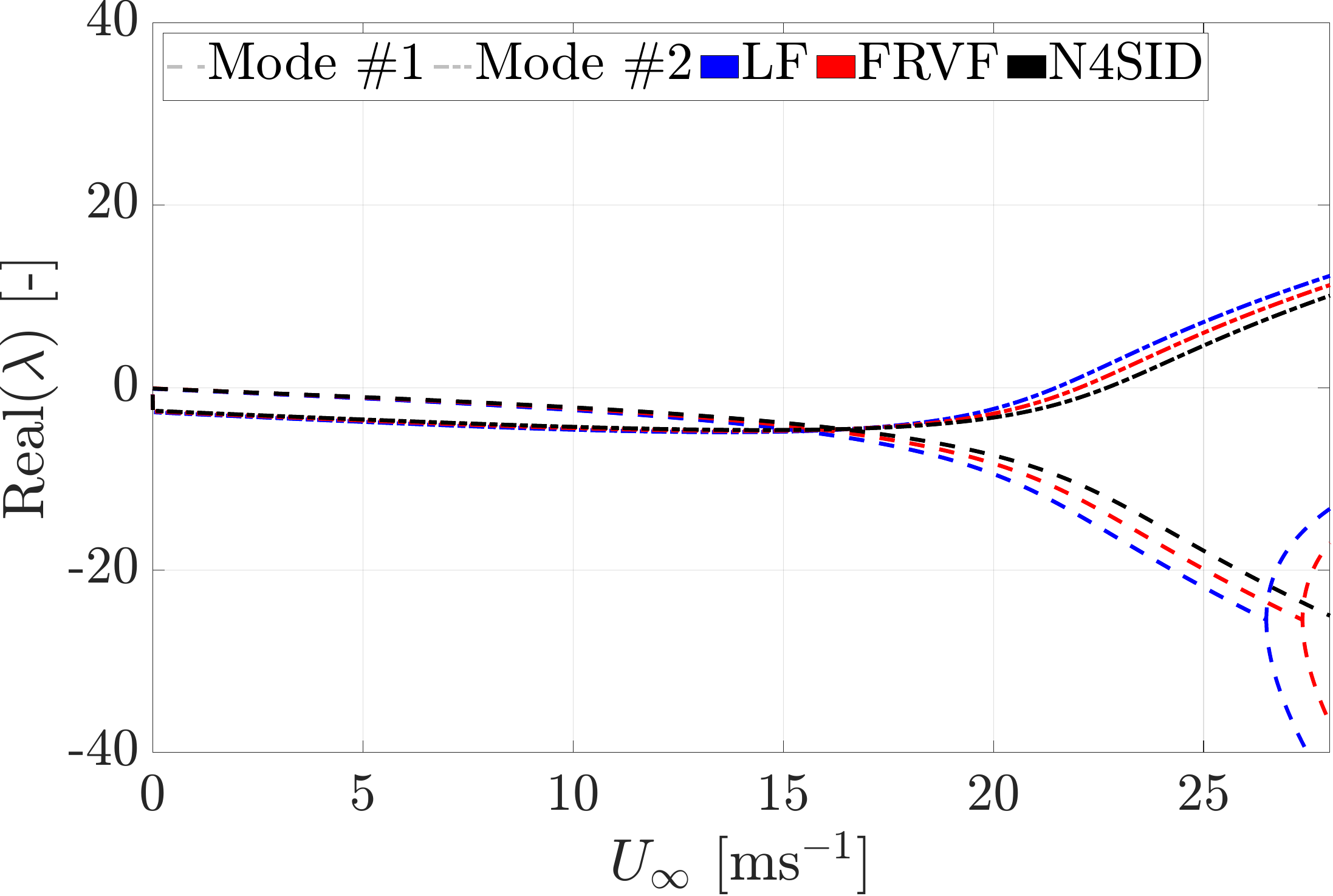}
      	\captionsetup{font={it},justification=centering}
		\subcaption{}
		\label{fig:flut1b}	
	\end{subfigure}   
    \vspace{1em}
	\begin{subfigure}{.49\textwidth}
		\centering
       \includegraphics[width=\textwidth]{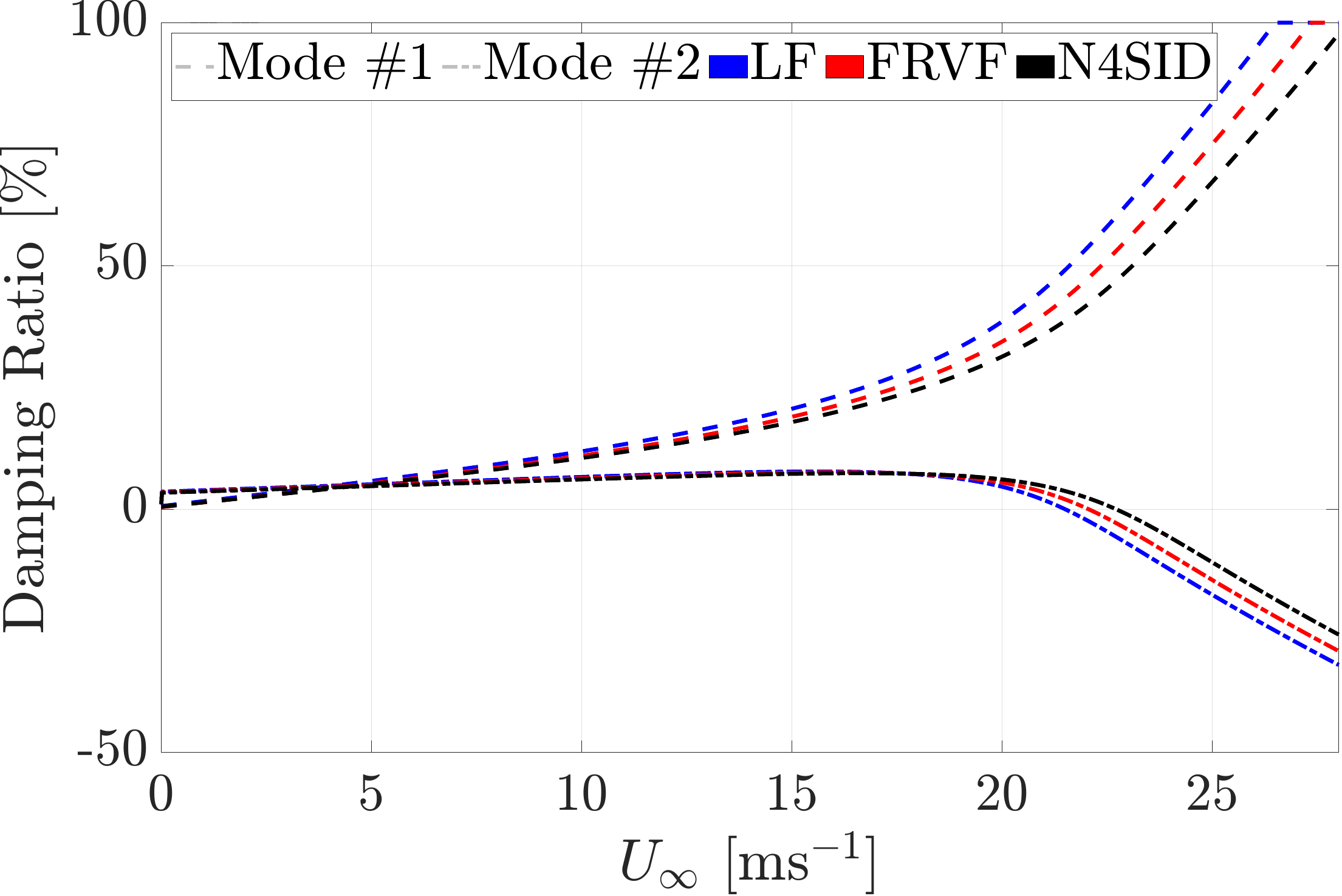}
      	\captionsetup{font={it},justification=centering}
		\subcaption{}
		\label{fig:flut1c}
   \end{subfigure}
   \begin{subfigure}{.49\textwidth}
		\centering
       \includegraphics[width=\textwidth]{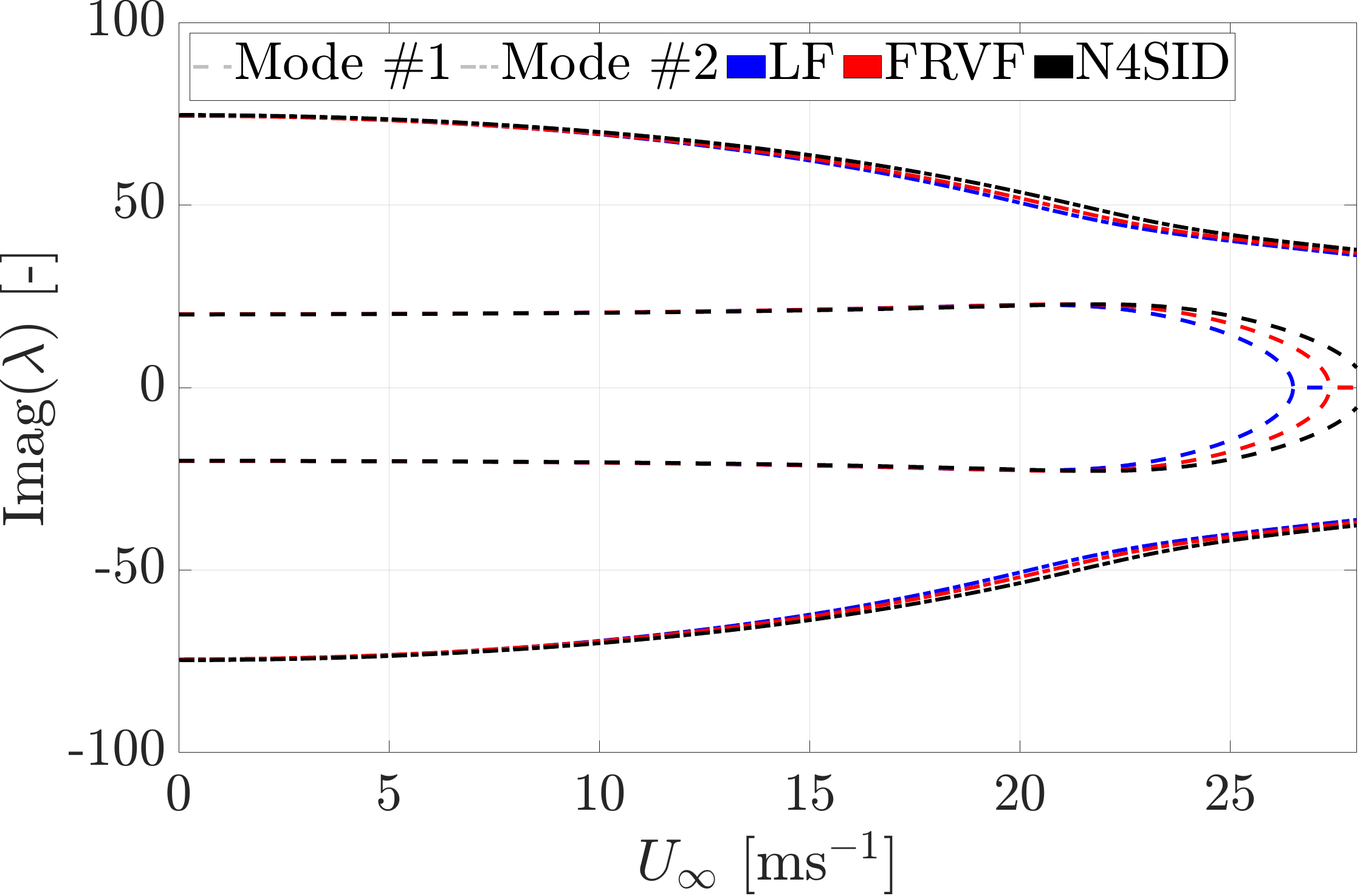}
      	\captionsetup{font={it},justification=centering}
		\subcaption{}
		\label{fig:flut1d}	 
	\end{subfigure}
	\caption{XB-2: Frequency (\cref{fig:flut1a}), damping ratio (\cref{fig:flut1c}){, real }(\cref{fig:flut1b}){ and imaginary part }(\cref{fig:flut1d}){ of the eigenvalues} vs $U_\infty$ plots for the aeroelastic model computed from data obtained via LF, FRVF, and N4SID for the baseline scenario.
 \label{fig:flut1}}
\end{figure}

\begin{figure}[!ht]
\centering
      \includegraphics[width=.7\textwidth]{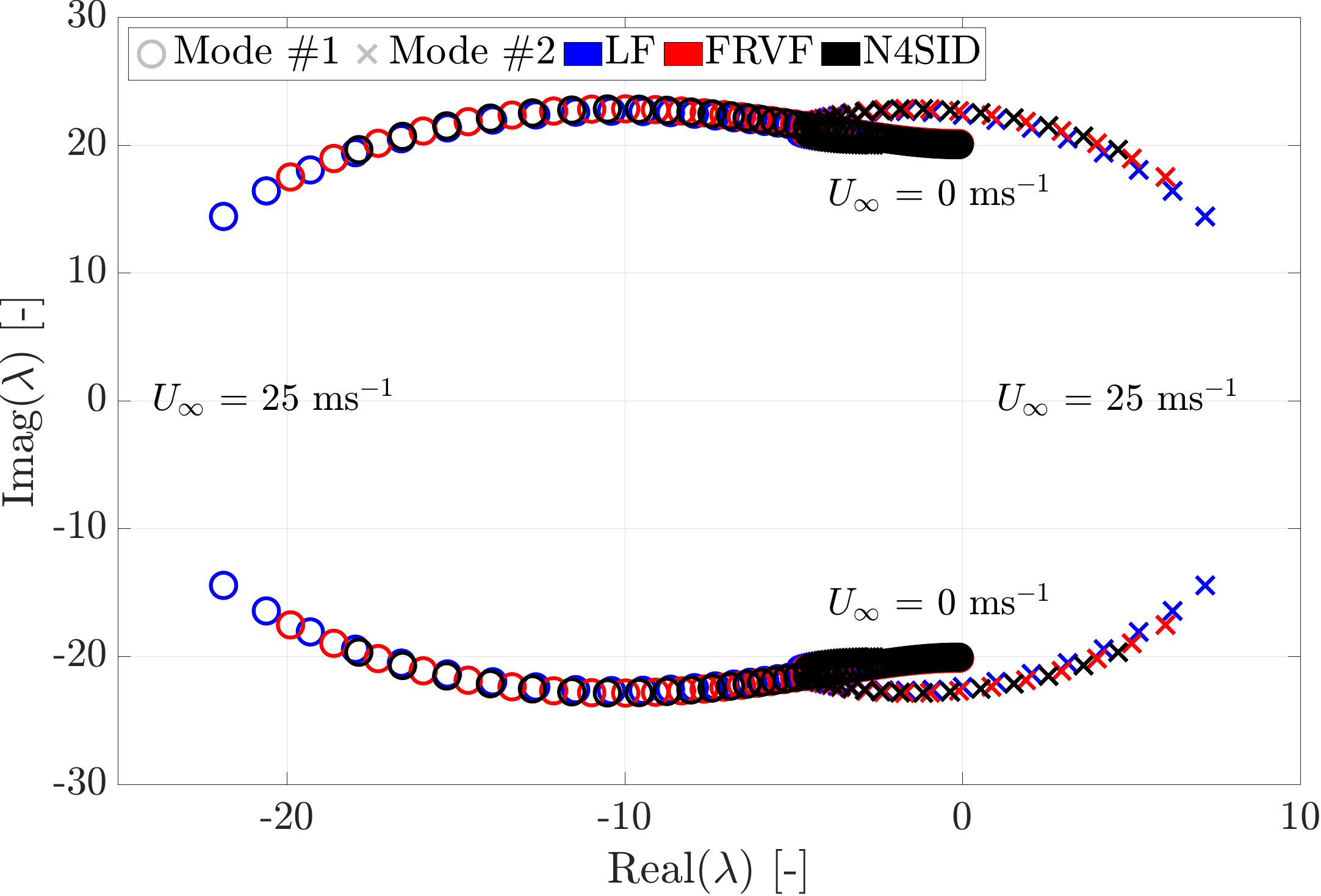}
	\caption{XB-2: Polar plot of the eigenvalues obtained from the LF, FRVF, and N4SID data for the baseline scenario.\label{fig:eiga}}
\end{figure}

\Cref{fig:flut1} shows the results, in terms of $\omega_n$, $\zeta_n$, and $\lambda$ vs $U_\infty$ of the linear eigenvalue analysis of the aeroelastic model, respectively from the LF, FRVF, and N4SID identified data, for the baseline case. In particular, \cref{fig:flut1a} shows results for $\omega_n$ vs $U_\infty$, \cref{fig:flut1c} plots the $\zeta_n$ vs $U_\infty$ curves. To get an insight into the stability of the $\lambda$ and confirm divergence does not proceed flutter, their real part is plotted in \cref{fig:eiga} against the respective imaginary part for the baseline scenario for the results obtained from LF, FRVF. and N4SID.

\begin{table}
   \centering
    \caption{{XB-2: Aeroelastic phenomena onset speeds and the difference relative to N4SID-derived values.}\label{tab:flutter}}
    \begin{tabular}{lcccc}
        \toprule
        \multicolumn{5}{c}{\textbf{Flutter onset speed }[ms\textsuperscript{-1}] (Difference wrt N4SID [\%])}\\\midrule
        \textbf{Scenario} & Baseline &  2 & 3 & 4 \\\hline
        \textbf{N4SID} & 22.710 & 23.336 & 23.285 & 23.205 \\
        \textbf{LF} & 21.498 & 22.200 & 22.116 & 21.991\\
        &  (-5.34) & (-4.87) & (-5.02) & (-5.23) \\
        \textbf{FRVF} & 22.057 & 22.743 & 22.727 & 22.590 \\
        &  (-2.88) & (-2.54) & (-2.40) & (-2.65) \\\bottomrule
    \end{tabular}
\end{table}

It should be considered that, in a real scenario, unless flutter is actively suppressed, the aircraft, or component therein, would be seriously damaged or fail at the flutter speed, making the definition of the divergence speed redundant.
The stability behaviour in \cref{fig:flut1} is similar to what is expected from similar implementations \citep{Afonso2017}. For the case reported in \cref{fig:flut1}, flutter, as expected, is related to $\zeta_n$ approaching zero. These phenomena can also be appreciated in \cref{fig:eiga}, where the zero line of the imaginary axis is crossed by $\lambda$, pointing to the occurrence of aeroelastic instability. {In addition, the absence of zero-ed imaginary values in} \cref{fig:eiga} {confirms that flutter preeceds divergence for the case under scrutiny.} These behaviours are also found in the cases for which plots are not presented.

The values for flutter onset speeds, reported in \cref{tab:flutter}, show that, when compared with the N4SID-derived models, the FRVF-derived models predicted onset speeds are much closer than those from the LF-derived model. Particularly, the difference between the FRVF-derived models and those from N4SID never exceeds {2.88}\%. However, even in the worst case, the error of the LF-derived model does not exceed {5.34}\%. The minimum deviations are also different; FRVF-derived models have a minimum error of {2.36}\%, while for LF models, it is {4.87}\%. Nevertheless, it should be considered that for the small mass changes implemented (less than 10\%), the flutter onset speed change is less than 5\%, wrt the baseline case for all methods.
It can be said that the LF modal identification performs worse than that via FRVF for the creation of a 2 DoF ROM for aeroelastic phenomena onset speed predictions. However, the LF-derived models are still able to translate the small changes in modal properties to changes in the predicted flutter speeds, {as the expected error is 20\% }\citep{keane2017}. 
In order to understand this, we need to take a look at \cref{tab:freq}. Now, the $\omega_n$ identified via LF and FRVF have negligible errors (<0.5 \%) wrt to the N4SID-values; however, this changes for $\zeta_n$, where the lowest, in absolute value, error is 4\%. For all cases, excluding $\zeta_1$ from scenario 2, the $\zeta_n$ identified via LF have a higher, albeit small, error than those from FRVF. This small error propagates in the aeroelastic phenomena onset speed computation, resulting, for the LF-derived values, in an absolute relative error double ($\approx$5\%) to that of the FRVF-derived speed.

{The reduced-order model adopted in this study is based on a number of simplifying assumptions that, while effective for preliminary assessments, limit the generality of the results. First, the wing is assumed to be rectangular, with constant flexural and torsional stiffness and uniform mass distribution. These assumptions neglect structural complexity such as spanwise variation in material properties and geometric discontinuities. Second, only the first flapping and first torsional modes are considered. While suitable for capturing primary flutter dynamics, this simplification excludes higher-order modes and potential modal couplings that may become significant in more flexible or complex wings. Third, the model employs a classical linear aeroelastic approach using the $p-k$ method, which presumes harmonic motion and small damping (small amplitudes assumption). Nonlinear aerodynamic effects, which are known to influence flutter onset in real configurations, are therefore not captured. Moreover, its applicability to full-scale aircraft structures should be approached with caution, as it does not account for Reynolds number effects or high-speed aeroelastic phenomena.
Since the experimental data were obtained for a sine sweep at a very slow chirp rate, in a controlled experimental environment, using high quality and sensitivity accelerometers, the influence of noise or other uncertainty has not been considered in this work. However, if a similar study is to be carried out using other-than-ideal experimental or operational scenarios, these should be considered.
Furthermore, the computational performance and scalability of LF, FRVF, and N4SID is not addressed in this work as it has already been considered in the literature. See} \cite{Dessena2022f, Dessena2022g} {for, respectively, studies on LF and N4SID, and LF FRVF, and N4SID scalability and computational performance.}

\section{Conclusions\label{sec:conc}}
This study established the value of two frequency-domain system identification methods, specifically Fast Relaxed Vector Fitting (FRVF) and the Loewner Framework (LF), for preliminary aeroelastic assessments. The main findings can be summarized as follows:

\begin{itemize}
    \item The flutter speed of the XB-2 wing is predicted using a two-degree-of-freedom reduced-order model, demonstrating reliable performance across scenarios;
    \item FRVF and LF reliably estimate natural frequencies and damping ratios;
    \item The flutter speeds predicted by the LF- and FRVF-derived models align closely with those obtained from N4SID, with deviations not exceeding 5\% for LF and 3\% for FRVF;
    \item The variation in the damping ratio induced by the methods significantly affects flutter speed predictions, emphasising the importance of accurate damping ratio identification.
\end{itemize}

In conclusion, the accuracy of the identified damping ratio is critical for precise flutter speed predictions. Nevertheless, the deviations between the FRVF- and LF-derived parameters are within the reduced-order model acceptable margin of error (20\%). This highlights their potential as computationally efficient and robust alternatives to traditional modal identification methods to support preliminary flutter calculations. {However, the applicability of the proposed methodology is currently limited to small-scale wings, such as those found in UAVs below 7~kg or scaled laboratory models. Extrapolating the approach to larger aircraft structures requires careful consideration of scale-dependent phenomena, particularly the influence of Reynolds number and aeroelastic coupling at higher airspeeds. Additionally, executing GVTs on small UAVs and models can result in disproportionate costs, where instrumentation, software licensing, and data acquisition may exceed the cost of fabricating the test article itself. These limitations highlight the need for future research into cost-effective testing procedures, the integration of higher-fidelity aerodynamic and structural models, and validation campaigns across a broader range of scales. Such efforts will help establish the generality and practical deployment of the proposed identification techniques within advanced aeroelastic design workflows.}

{\small
\section*{{\small Corresponding Author}}
\noindent Gabriele Dessena \orcidlink{0000-0001-7394-9303}\\
E-mail address: \href{mailto:gdessena@ing.uc3m.es}{GDessena@ing.uc3m.es}\\

\section*{{\small Author Contributions}}
{\small\noindent Conceptualisation, G.D.; methodology, G.D., M.C., and A.P; software, G.D.; validation, G.D., M.C., A.P., D.I.I., and J.F.W.; formal analysis, G.D.; investigation, G.D.; resources,  A.P., D.I.I., J.F.W. and L.Z.F.; data curation, G.D. and M.C.; writing---original draft preparation, G.D.; writing---review and editing, G.D., M.C., A.P., D.I.I., J.F.W., and L.Z.F.; visualisation, G.D., M.C., and D.I.I.; supervision, D.I.I., J.F.W., and L.Z.F.; funding acquisition, L.Z.F..}

\section*{{\small Declaration of conflicting interests}}
\noindent {\small The author(s) declared no potential conflicts of interest with respect to the research, authorship, and/or publication of this article.}

\section*{{\small Funding}}
\noindent {\small
The authors from Cranfield University disclosed receipt of the following financial support for the research, authorship, and/or publication of this article: This work was supported by the Engineering and Physical Sciences Research Council (EPSRC) [grant number 2277626].
The third author is supported by the \emph{Centro Nazionale per la Mobilità Sostenibile} (MOST -- Sustainable Mobility Center), Spoke 7 (Cooperative Connected and Automated Mobility and Smart Infrastructures), Work Package 4 (Resilience of Networks, Structural Health Monitoring and Asset Management).}}

\section*{{\small Acknowledgements}}
\noindent {\small
The authors wish to express their gratitude to the reviewers for their thoroughness and highly constructive feedback.}

\section*{{\small Data Availability Statement}\label{sec:6_data}}
{\small \noindent Data supporting this study ($p-k$ method MATLAB implementation) are openly available from the Zenodo Repository at \url{https://doi.org/10.5281/zenodo.15176140}. Furthermore, this study used existing authors' data made available under licence at \url{https://doi.org/10.5281/zenodo.11635814} and derived from the following resource available in the public domain: \cite{Dessena2022g}.}

\bibliographystyle{elsarticle-num}

\end{document}